\documentclass[journal,twocolumn]{IEEEtran}
\usepackage[utf8]{inputenc}

\usepackage{amssymb}
\usepackage{color,soul}
\usepackage[T1]{fontenc}
\usepackage{multicol}
\usepackage{multirow}
\usepackage{textcomp}
\usepackage{multirow}
\usepackage{bigdelim}
\usepackage{graphicx}
\usepackage{epstopdf}
\usepackage{amssymb}
\usepackage{amsmath}
\usepackage{psfrag}
\usepackage{amsthm}
\usepackage[tight,footnotesize]{subfigure}
\usepackage{algorithmic,algorithm}
\usepackage{mathrsfs}
\usepackage{framed}
\usepackage{color}
\usepackage{multirow,enumerate}
\usepackage{bigdelim}
\definecolor{shadecolor}{rgb}{0.92,0.92,0.92}
\usepackage{soul}
\usepackage{color}
\usepackage{cuted}
\usepackage{cite}

\theoremstyle{definition}

\newtheorem{theorem}{Theorem}

\newtheorem{lemma}{Lemma}

\newtheorem{remark}{Remark}
\newtheorem{problem}{Problem}

\makeatletter
\newcommand{\vast}{\bBigg@{3.2}}
\newcommand{\Vast}{\bBigg@{4.5}}

\makeatother

\begin{document}

\title{Convex Parameterization and Optimization for Robust Tracking of a Magnetically Levitated \\Planar Positioning System}

\author{Jun Ma,
        Zilong Cheng,
        Haiyue Zhu,
        Xiaocong Li,\\
        Masayoshi Tomizuka, \IEEEmembership{Life Fellow,~IEEE,} and
        Tong Heng Lee

\thanks{Jun Ma is with the Robotics and Autonomous Systems Thrust, The Hong Kong University of Science and Technology,
Guangzhou, China, and also with the Department of Electronic and
Computer Engineering, The Hong Kong University of Science and Technology,
Clear Water Bay, Kowloon, Hong Kong SAR, China (e-mail:
jun.ma@ust.hk).}
\thanks{Zilong Cheng and Tong Heng Lee are with the Department of Electrical and Computer Engineering, National University of Singapore, Singapore 117583 (e-mail:  zilongcheng@u.nus.edu; eleleeth@nus.edu.sg).}
\thanks{Haiyue Zhu is with the Singapore Institute of Manufacturing Technology, Singapore 138634 (e-mail: zhu\_haiyue@SIMTech.a-star.edu.sg).}
\thanks{Xiaocong Li is with the John A. Paulson School of Engineering and Applied Sciences, Harvard University, Cambridge, MA 02138 USA (e-mail: xiaocongli@seas.harvard.edu).}
\thanks{Masayoshi Tomizuka is with the Department of Mechanical Engineering, University of California, Berkeley, CA 94720 USA (e-mail: tomizuka@berkeley.edu).}
\thanks{This work has been submitted to the IEEE for possible publication.
	Copyright may be transferred without notice, after which this version may
	no longer be accessible.}
}

\maketitle

\begin{abstract}
	Magnetic levitation positioning technology has attracted considerable research efforts and dedicated attention due to its extremely attractive features. The technology offers high-precision, contactless, dust/lubricant-free, multi-axis, and large-stroke positioning. In this work, we focus on the accurate and smooth tracking problem of a multi-axis magnetically levitated (maglev) planar positioning system for a specific S-curve reference trajectory. The floating characteristics and the multi-axis coupling make accurate identification of the system dynamics difficult, which lead to a challenge to design a high performance control system. Here, the tracking task is achieved by a 2-Degree of Freedom (DoF) controller consisting of a feedforward controller and a robust stabilizing feedback controller with a prescribed sparsity pattern. The approach proposed in this paper utilizes the basis of an ${H}_\infty$ controller formulation and a suitably established convex inner approximation. Particularly, a subset of robust stabilizable controllers with prescribed structural constraints is characterized in the parameter space, and so thus the re-formulated convex optimization problem can be easily solved by several powerful numerical algorithms and solvers. With this approach, the robust stability of the overall system is ensured with a satisfactory system performance despite the presence of parametric uncertainties. Furthermore, experimental results clearly demonstrate the effectiveness of the proposed approach.
\end{abstract}


\section{Introduction}

\IEEEPARstart{M}{agnetic} levitation is a promising motion solution methodology to provide integrated bearing and actuation for a positioning system~~\cite{ZhuTIE2017TIE,Xu2018TIE,li2020data}, and it is characterized by the merits of high-precision, multi-axis, large-stroke, non-contact, frictionless, dust/lubricant-free, vacuum compatible and so on. As a result, such magnetically levitated (maglev) planar positioning systems are suitable for many applications with extremely high requirements. For example, in semiconductor manufacturing, the Extreme Ultraviolet (EUV) lithography procedure requires a vacuum environment and ultra-high motion precision in a large stroke. Also, for certain applications that the friction is a concern of high-accuracy tracking~\cite{yao2015adaptive}, the maglev technology can certainly avoid its undesirable effect. Furthermore, some manufacturing processes, such as pharmaceutical manufacturing, are sensitive to the contaminating dust generated by the traditional motion systems equipped with contact bearings, etc.; and thus the dust/lubricant-free, vacuum compatible feature of the maglev motion system is desirable.

For a typical precision motion system such as the maglev planar positioning stages, high-performance tracking of a specific reference profile is an essential objective pursued by motion control researchers and engineers, where stringent tracking performance is required to be ensured in terms of both accuracy and smoothness~\cite{zhu2019magnetically,yuan2019fast,ma2019parameter}. Consequently, various research efforts have been carried out to improve the tracking performance via, say, the model-based approaches~\cite{chen2014mu,ma2017integrated,ma2019robust} or the data-driven approaches~\cite{radac2013data,li2017data,li2018data}. However, in many of the existing available works, the tracking controller is designed and synthesized for essentially rather general situations. For instance, some of the works would consider the general full-frequency $H_{2}/H_{\infty}$ specifications; and also, the optimization in some of such works is not catered to a specific reference trajectory (even though such specificity may be present). It is pertinent to note that in numerous actual applications such as in various scanning processes, the motion profile is normally well designed for each particular process. Under these circumstances, such an attempted optimization controller design approach considering only general situations may not yield a resulting outcome specifically optimal for the given reference trajectory, and then the tracking performance could actually be downgraded. Therefore, an integrated tracking controller optimization approach instead with respect to a given reference is a promising solution to further improve the tracking performance.
Generally, an optimal control problem targeting at tracking a specific reference trajectory is challenging, because the augmentation of the reference dynamics and the system model leads to structural constraints in the controller. Consequently, some of the existing optimal control theories cannot be employed directly due to these constraints. Considering the linear quadratic regulator (LQR) problem as an example, the well-known Riccati equation is no longer effective when the controller is under structural constraints~\cite{geromel1982optimal}. Indeed, these constraints make the optimization problem more challenging than its unconstrained counterpart because of the non-convex characteristics~\cite{fazelnia2016convex}.

It is well known that a reliable optimization result usually needs an accurate system model as a prerequisite~\cite{xu2017continuous}. However, for the 6-Degree of Freedom (DoF) maglev planar positioning stage in this work, the floating characteristics and the multi-axis coupling make it tough to identify the plant model accurately. Therefore, it is of vital importance to develop a robust control approach to accommodate such parametric uncertainties~\cite{hu2009coordinated}.
Adaptive robust control has been researched extensively to reduce the effect of parametric variations through online parameter adaptation, and uncompensated uncertainties can be handled via certain robust control laws. For example, the neural network learning based adaptive robust controller has been designed to deal with the unexpected disturbances in the maglev stage motion control, but due to the black-box property, it is difficult to properly ensure the stability mathematically, and the computational effort is extremely large so that it cannot be widely applied to the existing industrial controllers. Similar works addressing this problem can also be found in~\cite{yang2012sliding,sun2019adaptive}.
To guarantee the robustness, an integrator-augmented linear quadratic control scheme is implemented on a 6-DoF maglev positioner~\cite{Silva2013}, and the learning adaptive robust control (LARC) is proposed in~\cite{hu2016performance} to improve the tracking performance. Another vital methodology in robust control is the ${H}_\infty$ control approach, which relates the system performance to the $\gamma$ disturbance attenuation level~\cite{doyle1989state}. Notably, the problem of finding the minimal disturbance attenuation level is typically defined as the optimal ${H}_\infty$ control problem~\cite{xie1992robust}. Unfortunately, the approach does not explicitly take the parametric uncertainties into consideration, and thus it may possibly lead to controllers that are not sufficiently robust; and then control failures may occur due to the existence of parametric uncertainties. To cater to the effect of parametric uncertainties and derive a stabilizing controller, the extended state observer (ESO) and the disturbance observer (DOB)  are very powerful tools in control applications~\cite{chen2015disturbance,yao2013adaptive,yao2017active,tan2019disturbance}. Particularly,~\cite{yao2013adaptive} presents an adaptive robust control method for DC motors with the ESO, and this approach accounts for not only the structured uncertainties (i.e., parametric uncertainties) but also the unstructured uncertainties (i.e., nonlinear friction, external disturbances,  unmodeled dynamics). Some remarkable results have also been reported in \cite{yao2017active} to handle both parametric uncertainties and disturbances, with the original contribution to invent a nonlinear interactive mechanism of disturbance observation and parametric adaptation. A disturbance compensation scheme is proposed in~\cite{tan2019disturbance}, which alters the reference profile and compensates for the disturbance using the DOB, such that improved tracking performance is attained in the tray indexing application. Also, considerable attention is commonly drawn towards the use of parameterization techniques, for example, the Youla-Kucera parameterization~\cite{Khatibi2010}, and alternative parameterization techniques based on the positive real lemma and the bounded real lemma~\cite{rantzer1994convex,ma2019robust2}. However, these methods still require various further developments and modifications before they can be accepted as effective solutions in the presence of parametric uncertainties and structural constraints, and especially also as solutions with attendant manageable numerical computation.

In this work, an effective robust control synthesis approach is developed for motion control of a maglev planar positioning stage, with the primary objective of achieving an accurate and smooth S-curve tracking task with suppressed control input chattering. This is developed and attained through a well-accepted structure of a 2-DoF controller architecture consisting of a feedforward controller and a feedback controller. Specifically, for the feedback controller design, an ${H}_\infty$ optimal control problem with a prescribed sparse-structured controller gain is formulated with the existence of parametric uncertainties. Through the construction of a convex parameterization (which can be considered as a variant of Youla-Kucera parameterization), all the feasible extreme models perturbed by the parametric uncertainties are explicitly considered in the parameter space, over which a set of robust stabilizing feedback controllers with the prescribed sparsity pattern are parameterized. Then, a resulting convex programming problem is formulated and efficiently solved.

The main contributions of this paper are listed below. With appropriate consideration of the tracking error signal and the feedback control input chattering, the prescribed accurate and smooth tracking task for the maglev planar positioning stage is formulated as an $H_\infty$ control related problem. A convex parameterization technique is presented (based on a variant of Youla-Kucera parameterization and convex restriction), such that the optimal ${H}_\infty$ control problem in the presence of parametric uncertainties is transformed into a convex optimization problem. This then enables the efficient determination of the feedback controller parameters for the maglev planar positioning stage. Furthermore, the designed controller ensures the required closed-loop stability and also maintains the system performance to be at the optimal level despite the presence of parametric uncertainties in the maglev planar positioning stage. Through a series of real-time experiments, the system performance of the maglev planar positioning stage is enhanced in terms of the accuracy and smoothness in the tracking problem.

The remainder of this paper is organized as follows. In Section II, system design and problem formulation of the tracking problem in the maglev planar positioning stage are given. To achieve the accurate and smooth S-curve tracking task, Section III presents the robust controller synthesis approach by convex parameterization. Next, in Section IV, to validate the proposed approach, optimization and experiments are conducted with detailed analysis. Finally, conclusions are drawn in Section V.

\section{System Description and Problem Formulation}

%

\begin{figure}[t]
	\centering
	\includegraphics[width=1\columnwidth]{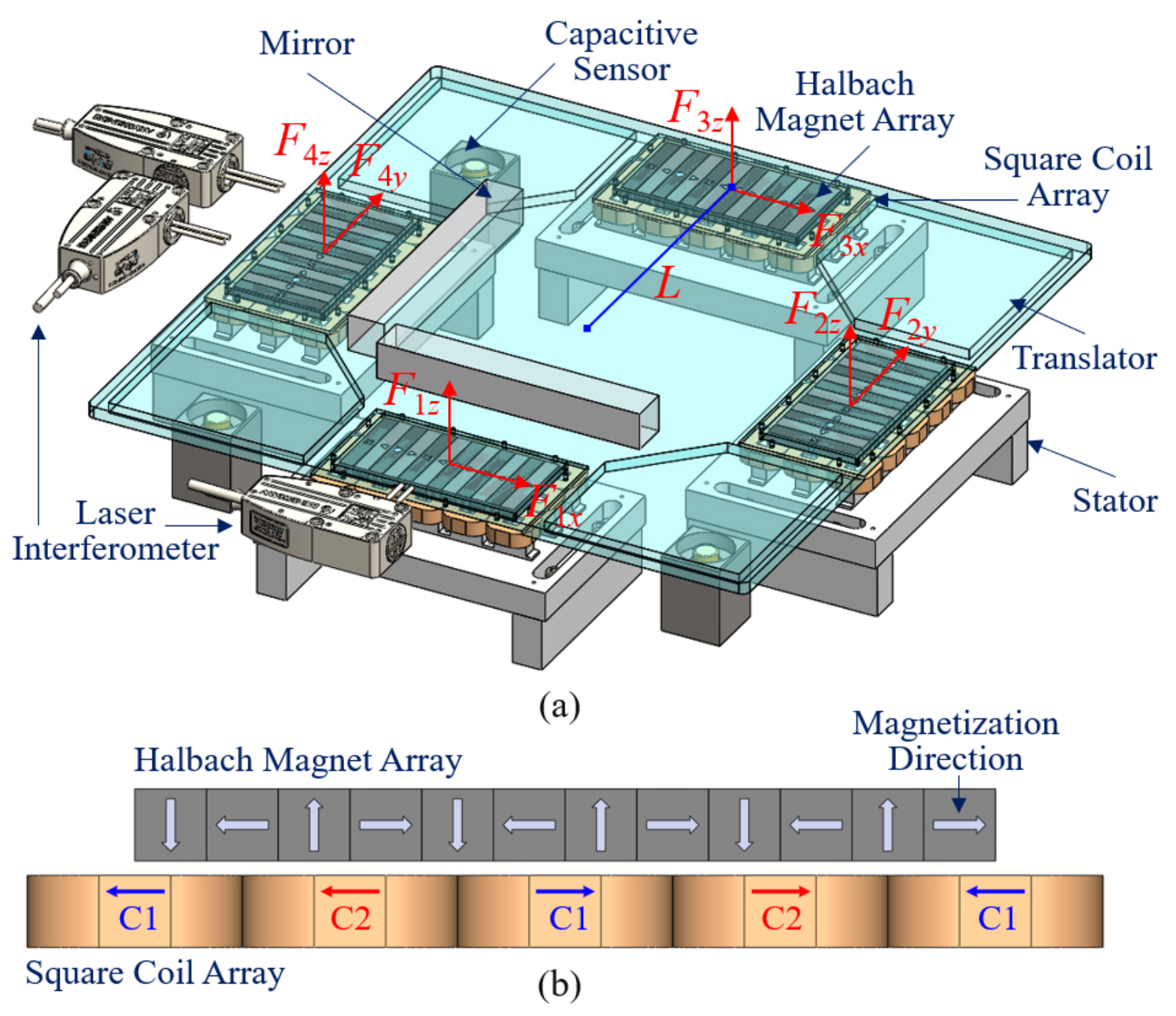}
	\caption{(a) Schematic illustration of the 6-DoF maglev planar positioning system in this work and (b) Working principle of the actuation in each forcer.}
	\label{fig:Schematics}
\end{figure}

\subsection{System Description}

The schematic illustration of the 6-DoF maglev planar positioning system is shown in Fig.~\ref{fig:Schematics}(a). Basically, the whole motion system consists of two parts, i.e. the stator and the translator. The stator is fixed on the ground, and it has four square coil arrays, which are placed crosswise on their platforms as shown in Fig.~\ref{fig:Schematics}(a). Each coil array has 15 square coils, which has a motion range of 30\,mm$\times$30\,mm with ideal force generation. The translator is the movable stage to deliver the 6-DoF positioning with centimeter-level planar motion, which is actually a panel carrying four Halbach magnet arrays. Each Halbach magnet array is placed on the top of the square coil array as shown in Fig.~\ref{fig:Schematics}(a), and each pair forms a forcer to provide both the controllable levitation force vertically and thrust force horizontally, where each of the forcers is shown in Fig.~\ref{fig:Schematics}(b). The coils in each forcer are grouped into two phases denoted by C1 and C2, where the adjunct two coils in each phase are injected with the same current magnitude but opposite directions.

In this work, the 6-DoF position feedback is provided by three channels of laser interferometers and three channels of capacitive sensors as indicated in Fig.~\ref{fig:Schematics}(a), and two reflective mirrors are installed together with the translator as the interferometer targets. As a result, the translator can be levitated during the motion without any physical contact with the stator. The 6-DoF motion of this maglev planar positioning system is controlled by the total levitation forces and thrust torques generated from the four forcers as mentioned above, which are determined by the 8-phase current inputs. The total forces and torques can be decoupled mathematically through the modelling for each channel of translation or rotation along $x-$, $y-$, and $z-$axes, where the detailed analysis on the system modelling can be found in~\cite{ZhuTIE2017TIE}.

It is worthwhile to mention that as detailed in~\cite{ZhuTIE2017TIE}, the 6-DoF maglev motion is a decoupled mechanism, and this is attained already and directly from the mechanism design level. As shown in Fig.~\ref{fig:Schematics}, in our 6-DoF maglev design, eight local forces ($F_{1x}$, $F_{2y}$, $F_{3x}$, $F_{4y}$ in the horizontal direction and $F_{1z}$, $F_{2z}$, $F_{3z}$, $F_{4z}$ in the vertical direction) are used to generate the 6-DoF force and torque for delivering the 6-DoF motion. Indeed, it is an over-actuated system and the 6-DoF global force and torque can be represented by $F_g=MF_l$, where $F_g=\begin{bmatrix} F_x & F_y & F_z & T_x & T_y & T_z \end{bmatrix}^T$, $F_l=\begin{bmatrix} F_{1x} & F_{1z} &F_{2y} &F_{2z} &F_{3x} &F_{3z} &F_{4y} &F_{4z} \end{bmatrix}^T$,
and
\begin{IEEEeqnarray}{rCl}
M=\begin{bmatrix}
0 & 0 & 1 & 0 & 0 & 0 & 1 & 0\\
1 & 0 & 0 & 0 & 1 & 0 & 0 & 0\\
0 & 1 & 0 & 1 & 0 & 1 & 0 & 1\\
0 & 0 & 0 & -L & 0 & 0 & 0 & L\\
0 & -L & 0 & 0 & 0 & L & 0 & 0\\
L & 0 & -L & 0 & -L & 0 & L & 0
\end{bmatrix}.
\end{IEEEeqnarray}
Note that $F_x$, $F_y$, and $F_z$ represent the global forces in $x-$, $y-$, and $z-$axes, respectively, $T_x$, $T_y$, and $T_z$ represent the global torque in $x-$, $y-$, and $z-$axes, respectively, and $L$ denotes the length from platform center to the magnet array. Using the decoupling mechanism $F_l = (M^T M)^{-1}M^T F_g$, the force and torque are decoupled and individually controlled for each channel of DoF. As the remaining coupling effects are not significant essentially, the 6-DoF maglev positioning system is considered as a 6-channel Single-Input-Single-Output (SISO) system in this work. In our approach, we treat the coupling effects passively by improving the controller's disturbance rejection capability to suppress the remaining coupling effects.


\subsection{Problem Formulation}

In this work, our control target focuses on the tracking performance of the planar motion. Individually, each axis of motion can be viewed as a SISO system. As there is no mechanical stiffness in the system (no mechanical bearing and contact), the plant model of the $x-$, $y-$axes motion without consideration of the external disturbance can be expressed as a standard second-order motion system given by
\begin{IEEEeqnarray}{rCl}~\label{eq:permodelnomimal}
m\ddot y +d\dot y  = u,
\end{IEEEeqnarray}
where $m$ and $d$ denote the lumped mass and damping for the motion system, $u$ and $y$ represent the control input and the system output (actual position) of the maglev planar positioning system, respectively. Note that although there is no friction, the eddy current force generated by the moving magnetic field is the important source of damping in~\eqref{eq:permodelnomimal}.

For the maglev positioning system, the rigid moving platform is directly driven by the controlled 6-DoF force and torque to deliver the 6-DoF motion, without any guiding mechanism or contacted support. As a result, it can be generally viewed as an ideal mass-damper system (no stiffness). Compared with other multi-DoF systems such as the flexure-based motion systems, there are much less unmodeled dynamics in the maglev motion system, because it doesn’t have the flexible structure that introduces the unmodeled dynamics or uncertainties. Therefore, it is reasonable to model the maglev system as a second-order system.

It is essentially useful to take parametric uncertainties into consideration for the proposed control problem. First of all, the floating characteristics and the multi-axis coupling make  accurate identification of the  system dynamics difficult, and undesired uncertainties in the measured data inevitably lead to challenges of deriving an extremely accurate system model.
Secondly, the maglev planar positioning system is always subjected to perturbations, such as applied loadings, levitation height variation, etc, which can also be formulated as parametric uncertainties.

Here, it is assumed that in the system~\eqref{eq:permodelnomimal}, $m$ and $d$ are subjected to parametric uncertainties $\Delta m$ and $\Delta d$, respectively, then it is generalized to a perturbed model characterized by
\begin{IEEEeqnarray}{rCl}~\label{eq:permodel}
(m+\Delta m)\ddot y +(d+\Delta d)\dot y = u ,
\end{IEEEeqnarray}
where $u$ and $y$ represent the control input and the actual position of the maglev planar positioning stage, respectively.

In this work, the maglev planar positioning stage is aimed at following a point-to-point reference profile accurately and smoothly. To design a smooth reference profile $r$, the following differential equation is proposed under certain initial conditions, where
\begin{IEEEeqnarray}{rCl}~\label{eq:frank}
\frac{d^n}{dt^n}r+z_1\frac{d^{n-1}}{dt^{n-1}}r+\cdots+z_{n} r=0,
\end{IEEEeqnarray}
for a given degree $n$ and a set of coefficients $z_i$, $\forall i=1,2,\cdots,n$.

Here we set $n=3$, and thus a third-order reference profile is implemented, which can be further expressed in the observability canonical form of the state-space representation:
\begin{IEEEeqnarray}{rcl}~\label{eq:ref}
\dot \rho=A_z \rho, \quad
r = C_z \rho,
\end{IEEEeqnarray}
with $\rho=\begin{bmatrix}
p & \dot p& \ddot p
\end{bmatrix}^T$, $A_z=\begin{bmatrix}
0& 1& 0 \\ 0&0& 1 \\z_1&z_2&z_3\end{bmatrix}$ and $C_z = \begin{bmatrix}1&0&0\end{bmatrix}$, thus it follows that $r=p$.

\begin{remark}
	The entries of the state matrix $A_z$ are zero except for a super-diagonal of 1's and the bottom row of coefficients, and \eqref{eq:ref} is referred to as a command-generator system. Also, due to its special shape, it is considered as an S-curve trajectory.
\end{remark}
\begin{remark}
	S-curve reference trajectories are widely used in motion control problems because of its smoothness, and a well designed S-curve trajectory can suppress residual vibrations induced to the system rather effectively.
Generally, the S-curve trajectory is predefined by the user depending on the specific industrial application (for different applications, the objective varies, because they may need different order of the S-curve trajectory with different speed of response in position, velocity, and acceleration profiles)~\cite{li2016integrated}.  
In our case, the reason that we set $n$ to be 3 is that there are 3 state variables in terms of the system ($e$, $\dot e$, and $\ddot e$) which would be used in the state vector (the state vector is to be defined later for constructing the PID control architecture). 
If a reference signal with an order of less than 3 is chosen, the reference jerk will not be bounded, which could induce residual vibrations to the system. 
If a reference signal with an order of higher than 3 is selected, extra state variables for the S-curve trajectory would be augmented in the state-space system, which are redundant for the tracking problem formulation and optimization. Therefore, we set $n$ to be 3. 
\end{remark}

Define the position error $e=r-y$, and then \eqref{eq:permodel} is expressed as
\begin{IEEEeqnarray}{rCl}~\label{eq:errormodel}
	(m+\Delta m)(\ddot r-\ddot e)+(d+\Delta d)(\dot r-\dot e) =u.
\end{IEEEeqnarray}
A 2-DoF control scheme is employed in this work, which comprises a feedforward controller $u_{ff}$ and a feedback controller $u_{fb}$, which means
\begin{IEEEeqnarray}{rCl}
	u=u_{ff}+u_{fb}.
\end{IEEEeqnarray}
	 Here, the feedforward controller is designed as \begin{IEEEeqnarray}{rCl}~\label{eq:ff} u_{ff}= m\ddot r+d\dot r,\end{IEEEeqnarray} and then \eqref{eq:errormodel} gives
\begin{IEEEeqnarray}{rcl}~\label{eq:errormodel2}
	(m+\Delta m)\ddot e+(d+\Delta d)\dot e-\Delta m \ddot r-\Delta d \dot r+u_{fb}=0.
\end{IEEEeqnarray}

Since smooth tracking is required in our problem, the feedback control input chattering needs to be suppressed to an adequate level with no strong vibration incurred in the experiment. To build a model that considers the feedback control input chattering as the fictitious control input, we take the derivative of \eqref{eq:errormodel2}, then
\begin{IEEEeqnarray}{rcl}\label{eq:errormodel3}
	(m+\Delta m)\dddot e+(d+\Delta d)\ddot e-\Delta m \dddot r-\Delta d \ddot r+\dot u_{fb}=0.
\end{IEEEeqnarray}
From the S-curve reference trajectory~\eqref{eq:ref}, we have
\begin{IEEEeqnarray}{rcl}\label{eq:refmodel}
\dddot r=\dddot p=z_1p+z_2 \dot p+z_3\ddot p.
\end{IEEEeqnarray}
Plug~\eqref{eq:refmodel} into \eqref{eq:errormodel3}, we have
\begin{IEEEeqnarray}{l}
(m+\Delta m)\dddot e+(d+\Delta d)\ddot e\IEEEnonumber\\
\quad\;-\Delta m (z_1p+z_2\dot p+z_3 \ddot p)-\Delta d \ddot p+\dot u_{fb}=0.
\end{IEEEeqnarray}

Define the state vector $x=\begin{bmatrix}
p & \dot p & \ddot p& e & \dot e&\ddot e
\end{bmatrix}^T$, and then we aim to build an augmented model by integrating the reference trajectory with the plant model. Assume the experiments are conducted under certain exogenous disturbances, then a disturbance vector $w$ is added into the augmented model, and thus the state-space representation of the system is given by
\begin{IEEEeqnarray}{rcl}\label{eq:sys}
	\dot x=Ax+B_2 \dot u_{fb}+ B_1 w,
\end{IEEEeqnarray}
	where
	\begin{IEEEeqnarray}{rCl}
		A=\left[\begin{array}{cccccc}
			0&1&0&0&0&0\\
			0&0&1&0&0&0\\
			z_1&z_2&z_3&0&0&0\\
			0&0&0&0&1&0\\
			0&0&0&0&0&1\\
			\frac{z_1\Delta m}{m+\Delta m}&\frac{z_2\Delta m}{m+\Delta m}&\frac{z_3\Delta m+\Delta d}{m+\Delta m}&0&0&-\frac{d+\Delta d}{m+\Delta m}
		\end{array}\right],\nonumber\IEEEeqnarraynumspace \end{IEEEeqnarray}
		\begin{IEEEeqnarray}{rCl}B_2&=&\left[\begin{array}{cccccc} 0&0&0&0&0&-\frac{1}{m+\Delta m}\end{array}\right]^T.\nonumber
	\end{IEEEeqnarray}
Remarkably, it is a common practice to assume $B_1$ to be a matrix with diagonal elements equal to 1 where necessary.

Since the state variables $p$, $\dot p$, and $\ddot p$ are not controlled (these state variables are only used to construct the augmented system), then the controlled output vector is defined as
\begin{IEEEeqnarray}{rcl}
z=Cx+D\dot u_{fb},
\end{IEEEeqnarray}
where $C=\begin{bmatrix} 0 &0 &0 &q_1 & 0 & 0\\ 0 &0 &0 &0 & q_2 & 0\\0 &0 &0 &0 & 0 & q_3\\ 0 &0 &0 &0 & 0 & 0\end{bmatrix}$ and $D=\begin{bmatrix} 0 \\ 0\\0\\r \end{bmatrix}$. The parameters $q_1$, $q_2$, $q_3$, and $r$ are considered as the weighting factors to be chosen according to the requirements.

The fictitious static state feedback controller is designed as 	
\begin{IEEEeqnarray}{rcl}~\label{eq:fic}
\dot u_{fb} = -Kx,
\end{IEEEeqnarray}
where the gain matrix is given by
\begin{IEEEeqnarray}{rcl}~\label{eq:gain}
K=\begin{bmatrix}
0 & 0 &0 &-k_i &-k_p & -k_d
\end{bmatrix},
\end{IEEEeqnarray}
such that \eqref{eq:fic} can be re-stored to the standard PID form
	\begin{IEEEeqnarray}{rcl}\label{zkk}
	u_{fb} = -K \int_0^t x \, d_\tau = k_p e +k_i\int_0^t e \, d_\tau +k_d \dot e,
	\end{IEEEeqnarray}
	 where $k_p$, $k_i$, and $k_d$ represent the proportional gain, integral gain, and derivative gain in the PID controller, respectively.
\begin{remark}
The proposed controller has a 2-DoF architecture. Particularly, \eqref{eq:ff} is a feedforward controller that is designed based on the reference trajectory and the nominal system model. Additionally, to make sure that the fictitious static state feedback controller~\eqref{eq:fic} can be restored to a PID controller in the implementation, $K$ is constrained to be a sparse matrix as in~\eqref{eq:gain}. Due to the existence of parametric uncertainties, the following development mainly focuses on the design of a robust stabilizing feedback controller~\eqref{zkk} with a prescribed sparsity pattern~\eqref{eq:gain} that contributes to the accuracy and smoothness in the reference tracking problem.
\end{remark}

\begin{remark}
 In regulation problems such as LQR, the system~\eqref{eq:sys} can be simplified to a model with a lower dimension. The reason is that there is no dynamics for the reference signal (where the reference is essentially a step function for such regulation problems), and thus it is straightforward to derive an error model for PID control through a state feedback framework. Noting such situations, it is pertinent then to be aware that for tracking problems (for example, with the S-curve signal as the reference), augmentation is then compulsory because one cannot derive an error state space model easily due to the dynamics of the reference.
In this case, involving the dynamic model of the reference signal has to be done for the optimization with respect to the a specific reference trajectory, and indeed, this augmentation would lead to extra structural constraints to the controller structure~\cite{ma2019parameter}, such as sparse constraint, equality constraint, etc.
\end{remark}

On the basis of the ${H}_\infty$ control formulation, we define a transfer function from $w$ to $z$, which is given by
\begin{IEEEeqnarray}{rCl}
	H(s)=(C- D K) (sI-A+ B_2 K)^{-1} B_1, \label{eq:transfer function}
\end{IEEEeqnarray}
with its ${H}_\infty$-norm defined as \begin{IEEEeqnarray}{rl}\|H(s)\|_\infty=\sup \limits_\omega \,\sigma_\textup{M}[H(j\omega)],\end{IEEEeqnarray} where  $\sigma_\textup{M}(\cdot)$ returns the maximum singular value. It is worth stating that the objective of the optimal ${H}_\infty$ control problem in the presence of parametric uncertainties is to keep the ${H}_\infty$-norm at the optimal level to ensure satisfying tracking performance, while maintaining good robustness at the same time, with the consideration of all feasible models.

\section{Robust Controller Synthesis with Convex Parameterization}

In the remaining text, $\mathbb R^{m\times n}$ ($\mathbb R^{n}$) denotes the real matrix with $m$ rows and $n$ columns ($n$ dimensional real column vector), $\mathbb S^{n}$ represents the $n$ dimensional real symmetric matrix, $\mathbb S^{n}_{+}$ ($\mathbb S^{n}_{++}$) denote the $n$ dimensional positive semi-definite (positive definite) real symmetric matrix. First, we consider the case without parametric uncertainty, then as an alternative representation of the open-loop system~\cite{geromel1996convex}, the following matrices are defined:
\begin{IEEEeqnarray}{rClrCl}
F&=&\begin{bmatrix}
A & -B_2\\
0_{1\times 6} & 0
\end{bmatrix},  &
\quad G&=&\begin{bmatrix}
0_{6\times 1}\\1
\end{bmatrix},  \nonumber\\
Q&=&\begin{bmatrix}
B_1 B_1^T &0_{6\times 1}\\0_{1\times 6}&0
\end{bmatrix} ,  & R&=&\begin{bmatrix}
C^T C &0_{6\times 1}\\0_{1\times 6}& D^T D
\end{bmatrix},
\end{IEEEeqnarray}
with
$
\Delta m = \Delta d = 0
$.
Then, define the matrix
\begin{IEEEeqnarray}{rl}
W=\begin{bmatrix}
W_1 & W_2 \\
W_2^T & W_3
\end{bmatrix} ,  \label{eqn:partition}
\end{IEEEeqnarray}
where $W_1 \in \mathbb{S}^{ 6  }_{++}$, $W_2 \in \mathbb{R}^{ 6}$, $W_3 \in \mathbb{R}$, and the matrical function \begin{IEEEeqnarray}{rl}\Theta(W, \mu)= FW+WF^T+WRW+\mu Q, \end{IEEEeqnarray} with $\mu=1/{\gamma}^2$, with $\gamma$ represents the disturbance attenuation level in the ${H}_\infty$ control formulation. Furthermore, $\Theta(W, \mu)$ is partitioned as \begin{IEEEeqnarray}{rl}~\label{eq:theta_mu}
\Theta(W, \mu)= \begin{bmatrix} \Theta_{1}(W, \mu) & \Theta_{2}(W) \\ \Theta_{2}^T(W) & \Theta_{3}(W) \end{bmatrix},
\end{IEEEeqnarray}
with $\Theta_{1}(W, \mu) \in \mathbb{S}^{ 6 }, \Theta_{2}(W) \in \mathbb{R}^{ 6 }, \Theta_{3}(W) \in \mathbb{R}$.

The following theorem defines a feasible set in the parameter space, and the mapping between $W$ and $K$ is constructed, which exhibits some important properties on the stability and performance guarantee of the closed-loop system.
\begin{theorem} \label{thm:convex}
	Define the set $\mathscr{C} = \{(W, \mu) \in \mathbb{S}^7\times \mathbb{R}: W  \succeq 0, \Theta_1(W, \mu) \preceq 0, \mu > 0$\}. Then the following statements hold:
	\begin{enumerate}[(a)]	
		\item Any $(W, \mu) \in \mathscr{C}$ generates a stabilizing gain $K=W_2^T W_1^{-1}$ that guarantees $\|H(s)\|_\infty \leq \gamma$ with $\gamma=1/\sqrt \mu>0$.
		\item 	At optimality, $(W^*,\mu^*)=\textup{argmax}\{\mu:(W, \mu)\in \mathscr{C}\}$ gives the optimal solution to the optimal ${H}_\infty$ control problem, with $K^*={W_2^*}^{T} {W_1^*}^{-1}$ and $\|H(s)\|_\infty^*=\gamma^*=1/\sqrt{\mu^*}$.
	\end{enumerate}
\end{theorem}

\noindent{\textbf{Proof of Theorem \ref{thm:convex}:}}
To prove Statement (a), the following lemma is used.
\begin{lemma}~\cite{scherer1989h}\label{lemma:Riccati}
	Given $\gamma >0$, if the closed-loop is observable, it is asymptotically stable and $\|H(s)\|_\infty \leq \gamma$ if and only if the Riccati inequality
	\begin{IEEEeqnarray}{l}
	(A-B_2 K)^T P+P 	(A-B_2 K)  \IEEEnonumber\\
	\quad\;\, +\gamma^{-2} P B_1 B_1^T P + (C-DK)^T  (C-DK) \preceq 0,
	\end{IEEEeqnarray}
	has a symmetric positive definite solution $P=P^T\succ 0$.
\end{lemma}

From Lemma \ref{lemma:Riccati}, there exists a symmetric positive definite solution $P=P^T\succ 0$ such that
\begin{IEEEeqnarray}{l}
(A-B_2 K)^T P +P (A-B_2 K)    \nonumber\\
\quad\;\,+ \mu P B_1 B_1^T P+ C^T C+K^T D^T DK \preceq 0. \label{lemma:Riccatiexample}
\end{IEEEeqnarray}
Since $P$ is nonsingular, we have
\begin{IEEEeqnarray}{l}
P^{-1} (A-B_2 K)^T   +  (A-B_2 K) P^{-1}    + \mu   B_1 B_1^T \nonumber\\
\quad\; + P^{-1} C^T C P^{-1} +P^{-1} K^T D^T DK P^{-1} \succ 0. \label{lemma:Riccatiexample2}
\end{IEEEeqnarray}
Denote $W_p=P^{-1}$, \eqref{lemma:Riccatiexample2} is equivalent to
\begin{IEEEeqnarray}{l}
(A-B_2 K) W_p+W_{p}(A-B_2 K)^T +W_{p}C^TCW_{p}\nonumber\\
\quad\;\,+W_{p}K^TD^TDKW_{p}  +\mu B_1B_1^T \preceq 0.
\end{IEEEeqnarray}
From~\eqref{eq:theta_mu}, we have
\begin{IEEEeqnarray}{rCl}
\Theta_{1}(W,\mu)&=&AW_1-B_2W_2^T+W_1A^T-W_2B_2^T\nonumber\\&&+W_1C^TCW_1+W_2D^TDW_2 +\mu B_1B_1^T. \IEEEeqnarraynumspace \label{eq:theta1}
\end{IEEEeqnarray}
Then, by setting $W_1=W_p$ and $W_2^T=KW_p$, then we have $K=W_2^T W_1^{-1}$ and $\Theta_{1}(W,\mu)\preceq 0.$
Therefore, we can construct $W= \begin{bmatrix} W_1 & W_1 K^T\\KW_1 &W_3 \end{bmatrix}$.  By Schur's complement, we can ensure $W \succeq 0$ by choosing $W_3\succeq K W_1 K^T$, which provides a norm bound for the gain matrix. Hence, $K=W_2^T W_1^{-1}$ is a stabilizing feedback controller gain generated from $(W, \mu) \in \mathscr{C}$, and it follows from Lemma \ref{lemma:Riccati} that $\|H(s)\|_\infty \leq \gamma$ is guaranteed. Statement (b) is direct consequence of Statement (a).  \hfill{\qed}

Theorem~\ref{thm:convex} determines the optimal ${H}_\infty$ controller gain $K$ without any additional structural constraint or parametric uncertainties. However, as indicated in~\eqref{eq:gain}, $K$ is a sparse matrix with three zero elements, certain structural constraints need to be imposed on $W$ such that the resulting $K$ satisfies the sparsity requirement, thus Theorem~\ref{thm:sparse} is presented, and then Theorem~\ref{thm:robust} extends the results to be applicable to uncertain systems.

\begin{theorem}~\label{thm:sparse}
		Define the set $\mathscr{C}_S = \{(W, \mu)\in \mathbb{S}^7\times \mathbb{R}: W  \succeq 0, \mathcal L (W) = 0, \Theta_1(W, \mu) \preceq 0, \mu > 0$\}, where $\mathcal L:\mathbb R^{7\times 7}\rightarrow \mathbb R^{3\times 4}$ defines a linear operator $\mathcal L(W)=V_1WV_2$ with $V_1 = \begin{bmatrix}
		I_3 & 0_{3\times 4}
		\end{bmatrix}$ and $V_2=\begin{bmatrix}
		0_{3 \times 4} \\ I_4
		\end{bmatrix}$, then the following statements hold:
	\begin{enumerate}[(a)]	
		\item Any $(W, \mu) \in \mathscr{C}_S$ generates a stabilizing gain $K=W_2^T W_1^{-1}$ with the sparsity pattern as in~\eqref{eq:gain} that guarantees $\|H(s)\|_\infty \leq \gamma$ with $\gamma=1/\sqrt \mu>0$.
		\item 	At optimality, $(W^*,\mu^*)=\textup{argmax}\{\mu:(W, \mu)\in \mathscr{C}_S\}$ gives the optimal solution to the sparse optimal ${H}_\infty$ control problem, with $K^*={W_2^*}^{T} {W_1^*}^{-1}$ and $\|H(s)\|_\infty^*=\gamma^*=1/\sqrt{\mu^*}$.
	\end{enumerate}
\end{theorem}

\noindent \textbf{Proof of Theorem~\ref{thm:sparse}:}
For Statement (a), $W_1$ and $W_2$ are split as
\begin{IEEEeqnarray}{rCl}
W_1 =  \begin{bmatrix}
W_{1,1} & W_{1,2}\\
W_{1,2}^T & W_{1,3}
\end{bmatrix},\quad
W_2 = \begin{bmatrix}
W_{2,1} \\
W_{2,2}
\end{bmatrix},
\end{IEEEeqnarray}
where $W_{1,1}\in \mathbb{S}^{3}$, $W_{1,2} \in \mathbb{R}^{3 \times 3}$, $W_{1,3} \in \mathbb{S}^{3}$, $W_{2,1} \in \mathbb{R}^{3}$, $W_{2,2} \in \mathbb{R}^{3}$.
Then,
\begin{IEEEeqnarray}{rCl}
\mathcal{L}(W)&=& \begin{bmatrix}
	I_3 & 0_{3\times 4}
\end{bmatrix}\begin{bmatrix}
W_{1,1} & W_{1,2} &W_{2,1}\\
W_{1,2}^T & W_{1,3} &W_{2,2}\\
W_{2,1}^T & W_{2,2}^T &W_{3}
\end{bmatrix}\begin{bmatrix}
0_{3 \times 4} \\ I_4
\end{bmatrix}\IEEEnonumber\\
&=&\begin{bmatrix}
W_{1,2} &W_{2,1}
\end{bmatrix}.
\end{IEEEeqnarray}
So $\mathcal{L}(W)=0$ implies $W_{1,2}=0$ and $W_{2,1}=0$. Then,
\begin{IEEEeqnarray}{rCl}
K
&=&\begin{bmatrix} 0_{1 \times 3} & W_{2,2}^T \end{bmatrix}  \begin{bmatrix}
W_{1,1}^{-1} & 0_{3 \times 3}\IEEEnonumber\\
 0_{3 \times 3} & W_{1,3}^{-1}
\end{bmatrix}   = \begin{bmatrix}
0_{1 \times 3}  &  W_{2,2}^T W_{1,3}^{-1}
\end{bmatrix}.
\end{IEEEeqnarray}
Since $W_{1,1}$ and $W_{1,3}$ are non-singular, thus the prescribed sparsity pattern of $K$ is ensured as shown in~\eqref{eq:gain}. Statement (b) proceeds along the same proof as Theorem~\ref{thm:convex}(b).  \hfill{\qed}

Then, it suffices to extend the above results to uncertain systems, where $\Delta m \neq 0$ and $\Delta d \neq 0$. For convex-bounded parametric uncertainties, $F$ belongs to a polyhedral domain which can be expressed as a convex combination of the extreme matrices $F_i$, then \begin{IEEEeqnarray}{rCl}~\label{eq:poly} F=\sum \limits_{i=1}^4 \xi_i F_i, \end{IEEEeqnarray} with  $
	F_i=\begin{bmatrix}
		A_i & -B_{2i}\\
		0_{1\times 6} & 0
	\end{bmatrix}$,
$	\xi_i \geq 0 $, $ \sum \limits_{i=1}^4   \xi_i=1 $.  Then, for $i=1,2,\ldots,4$, define  \begin{IEEEeqnarray}{rCl}\Theta_i(W,\mu)=F_i W+WF_i^T+WRW+\mu Q, \end{IEEEeqnarray}   and further partition it as
\begin{IEEEeqnarray}{rCl} \Theta_i(W,\mu) = \begin{bmatrix} \Theta_{1i}(W,\mu) & \Theta_{2i}(W) \\
\Theta_{2i}^T(W) & \Theta_{3i}(W) \end{bmatrix},\end{IEEEeqnarray}
with $\Theta_{1i}(W,\mu) \in \mathbb{S}^{6 },
\Theta_{2i}(W) \in \mathbb{R}^{ 6  },
\Theta_{3i}(W) \in \mathbb{R}$. Consequently, a mapping between $W$ and $K$ with the prescribed sparsity pattern can be constructed in the presence of parametric uncertainties, and the results are shown in Theorem~\ref{thm:robust}.

\begin{theorem} \label{thm:robust}
	Define the set $\mathscr{C}_{Ri} = \{(W, \mu)\in \mathbb{S}^7\times \mathbb{R}: W  \succeq 0, \mathcal L (W) = 0, \Theta_{1i}(W, \mu) \preceq 0, \forall i=1,2,\cdots,4, \mu > 0$\}, and $\mathscr{C}_{R}=\bigcap \limits_{i=1}^{4} \mathscr{C}_{Ri}$, then the following statements hold:
	\begin{enumerate}[(a)]	
		\item Any $(W, \mu) \in \mathscr{C}_R$ generates a robust stabilizing gain $K=W_2^T W_1^{-1}$ with the sparsity pattern as in~\eqref{eq:gain} that guarantees $\|H_i(s)\|_\infty \leq \gamma$ with $\gamma=1/\sqrt \mu>0$ under convex-bounded parametric uncertainties, where $\|H_i(s)\|_\infty$ represents the $H_\infty$-norm of the $i$th extreme system.
		\item At optimality, $(W^*,\mu^*)=\textup{argmax}\{\mu:(W, \mu)\in \mathscr{C}_R\}$ gives the optimal solution to the sparse optimal ${H}_\infty$ control problem in the presence of parametric uncertainties, with $K^*={W_2^*}^{T} {W_1^*}^{-1}$ and $\gamma^*=1/\sqrt{\mu^*}$.
	\end{enumerate}
\end{theorem}

\noindent{\textbf{Proof of Theorem \ref{thm:robust}:}}
The proof is straightforward as it is an extension of Theorem~\ref{thm:convex} and Theorem~\ref{thm:sparse}. \hfill{\qed}
 
Essentially, Theorem~\ref{thm:convex} considers the case without considering imposed structural constraints and parametric uncertainties, and Theorem~\ref{thm:sparse} addresses the structural constraints. Furthermore, Theorem \ref{thm:robust} extends these results by considering both the structural constraints and parametric uncertainties. By defining the set $\mathscr{C}_{Ri}$, all the extreme systems are suitably taken into consideration. Additionally, it is notable that the robust stability for the entire uncertain domain is checked at only the vertices of the convex polyhedron as defined in~\eqref{eq:poly}. On the other hand, for the robust performance, $\gamma^*$ is the upper bound to ${H}_\infty$-norm by taking all the extreme models into consideration. Thus if $\gamma^*$ is optimal, the system performance is attained at optimality in the presence of parametric uncertainties.

Based on the development and analysis above, the following optimization problem is formulated.
\begin{problem}\label{problem}
\begin{IEEEeqnarray}{rl}~\label{eq:opt1}
\displaystyle\operatorname*{maximize}_{(W,\mu)\in\mathbb S^{7}\times \mathbb R}
&\quad  \mu \nonumber\\
\operatorname*{subject\ to}&\quad
W \succeq 0\nonumber\\
&\quad \mathcal{L}(W)=0 \nonumber\\
&\quad \Theta_{1i}(W,\mu)   \preceq 0, \, \forall i=1,2,\cdots, 4  \nonumber \\
&\quad\mu>0.
\end{IEEEeqnarray}
\end{problem}

\begin{remark}
Because of the convexity of Problem 1, several efficient numerical algorithms and solvers can be used to obtain the global optimum, for example, interior-point method~\cite{helmberg1996interior}, augmented Lagrangian method (ALM)~\cite{ma2020optimal}, etc.
\end{remark}

\section{Experiment Validation}
%

\subsection{Optimization Implementation and Results}

In order to conduct the optimization, the nominal model of the maglev planar positioning stage is identified, where the system identification is carried out in $x$ axis by letting the $x$-axis motion track the sweeping-frequency sinusoidal signals with the frequency ranging from 1\,Hz to 1000\,Hz. For each point in the frequency range, the frequency response from the reference to the system output can be obtained respectively by using the Fast Fourier Transform (FFT). The identification is conducted in the case where the levitation height is 1\,mm, and finally, a second-order system is fitted and given as
$
	P(s)={1}/{(ms^2+ds)}
$,
where $m=1/400$, $d=1/200$.

\begin{figure}[t]
	\centering
	\includegraphics[trim=50 50 50 100,width=1\columnwidth]{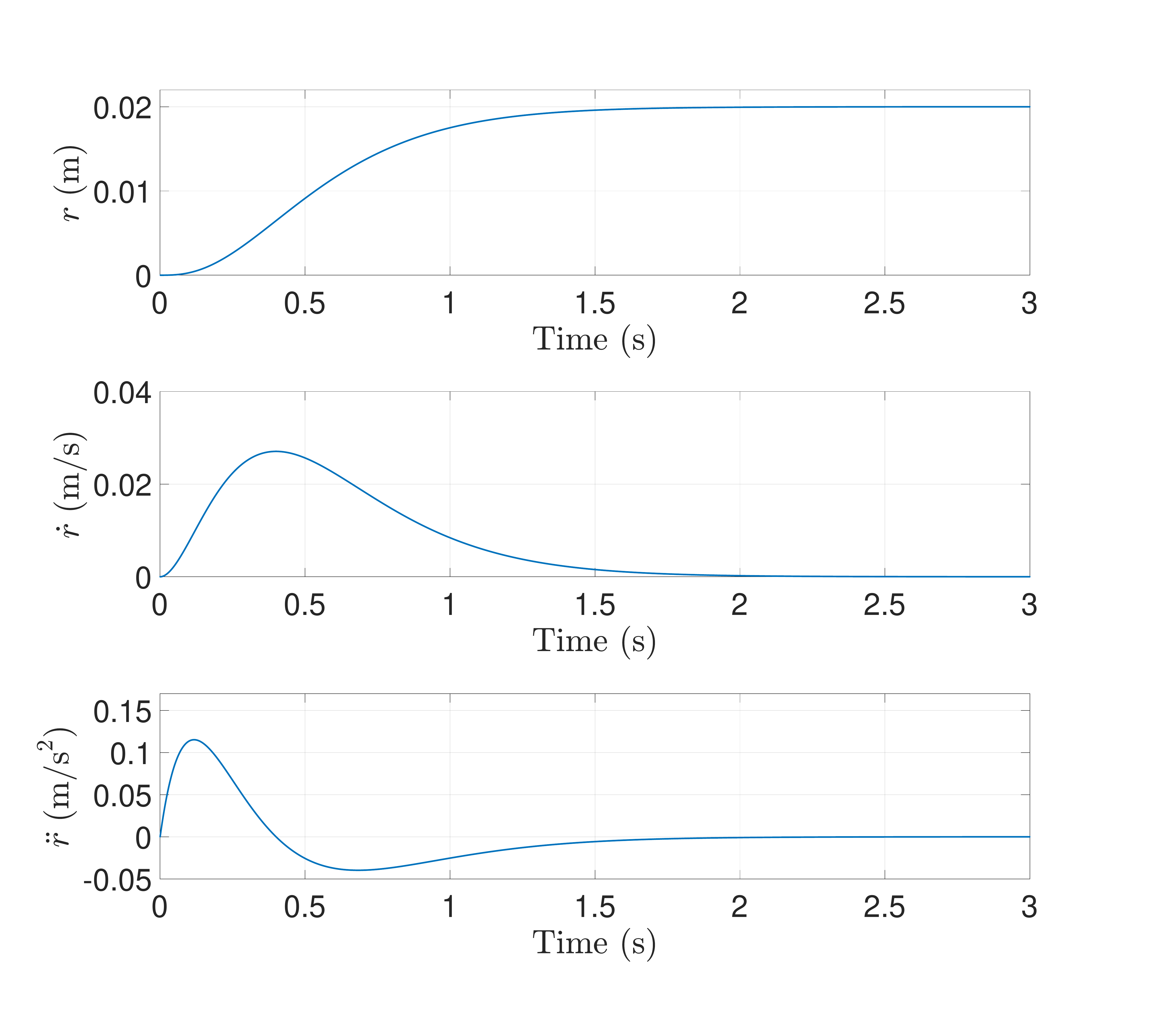}
	\caption{Reference profile used in the experiment.}
	\label{fig:ref}
\end{figure}

To solve the robust sparse optimization problem as formulated in the last section, many efficient optimization methods can be implemented. Furthermore, as Problem~\ref{problem} is a convex optimization problem, the global optimum can be obtained. However, note that there is a quadratic term $WRW$ included in the constraint $\Theta_{1i}(W,\mu)   \preceq 0, \, \forall i=1,2,\cdots, 4$, which impedes the solving of Problem~\ref{problem}, because most of the numerical algorithms and solvers are not capable of handling the quadratic constraints. Hence, the Schur complement is used to derive an equivalent expression of the constraint $\Theta_{1i}(W,\mu)   \preceq 0, \, \forall i=1,2,\cdots, 4$, such that the quadratic term can be transformed into the linear conic form. Thus the constraint $\Theta_{1i}(W,\mu)   \preceq 0, \, \forall i=1,2,\cdots, 4$ in Problem~\ref{problem} is equivalently denoted by
\begin{IEEEeqnarray}{rl}~\label{eq:schur}
\begin{bmatrix}
-VF_i WV^T-VW  F_i^TV^T-\mu VQV^T & VWR^{\frac{1}{2}}\\R^{\frac{1}{2}}WV^T & I_7
\end{bmatrix}  \succeq 0,  \nonumber\\
\end{IEEEeqnarray}
where $V = \begin{bmatrix} I_6&0_{6\times 1}\end{bmatrix}$.
In summary, Problem~\ref{problem2} is given below, which serves as an equivalent expression of Problem~\ref{problem}.
\begin{problem}\label{problem2}
	\begin{IEEEeqnarray}{rl}~\label{eq:opt2}
	\displaystyle\operatorname*{maximize}_{(W,\mu)\in\mathbb S^{7}\times \mathbb R}
	&\quad  \mu \nonumber\\
	\operatorname*{subject\ to}&\quad
	W \succeq 0\nonumber\\
	&\quad \mathcal{L}(W)=0 \nonumber\\
	&\quad\begin{bmatrix}
	-VF_i\\R^{\frac{1}{2}}
	\end{bmatrix}
	W
	\begin{bmatrix}
	V^T & 0_{7 \times 7}
	\end{bmatrix}\nonumber\\
	&\quad+\begin{bmatrix}V\\0_{7 \times 7}\end{bmatrix}W \begin{bmatrix}-F_i^TV^T &R^{\frac{1}{2}} \end{bmatrix} \nonumber\\
	&\quad+\mu  \begin{bmatrix}- VQV^T&0_{6 \times 7}\\0_{7 \times 6}&0_{7 \times 7}\end{bmatrix}  +  \begin{bmatrix} 0_{6 \times 6} &0_{6 \times 7}
	\\  0_{7 \times 6}&I_7 \end{bmatrix} \succeq 0  \nonumber\\
	&\quad\quad\quad\quad\quad\quad\quad\quad\quad\quad\quad\quad\quad\forall i=1,2,\cdots, 4  \nonumber \\
	&\quad\mu>0.
	\end{IEEEeqnarray}
\end{problem}
In this work, the parametric uncertainties are assumed such that $-30\% m \leq \Delta m \leq 30\% m$ and  $-30\% d \leq \Delta d \leq 30\% d$.
Also, the weighting parameters are given as $q_1=10^4$, $q_2=10^2$, $q_3=0$, and $r=1$. Notice that $q_3$ is set to be zero because our primitive objective is to ensure the accurate tracking ($q_1$ penalizes $e$) and smooth tracking ($q_2$ penalizes $\dot e$ and $r$ penalizes $\dot u_{fb}$).
As a standard practice, the weighting factors are usually user-defined in view of different control objectives and scenarios. The user is always able to make further adjustments if the optimization results based on these weighting parameters do not meet the specific requirement.
The parameters of the S-curve reference trajectory are chosen as $z_1 = -125$, $z_2 = -75$, and $z_3 = -15$, and the initial condition of the state vector is defined as $\rho_0=\begin{bmatrix}
-0.02\,\textup{mm} & 0\,\textup{mm/s}  & 0\,\textup{mm}/\textup{s}^2
\end{bmatrix}^T$. Then, the S-curve reference trajectory is shifted up for $0.02\,\textup{mm}$ such that the reference position starts from zero and the set-point is $0.02\,\textup{mm}$, as shown in Fig.~\ref{fig:ref}. Note that the result of the optimal controller is independent of the initial condition of the S-curve reference trajectory~\cite{ma2017integrated}.  Furthermore, since the order of the reference profile is three, the reference position, velocity, and acceleration are all bounded.

The optimization is implemented on a desktop with the CPU Intel(R) Xeon(R) CPU E5-2695 v3 @ 2.30GHz, based on the python optimization package CVXPY with the CVXOPT solver. At optimality, $W^*$ and $\mu^*$ are obtained, where $W^{*}$ is shown in~\eqref{eq:W} and $\mu^{*} = 1.0764\times 10^{-5}$. Then, $\gamma^*=304.7995$, and the optimal controller gain matrix $K^*$ is given by
\begin{IEEEeqnarray}{rCl}~\label{eq:optgain}
K^* = \begin{bmatrix}
0&
0&
0&
-1664.71&
-47.71&
-0.50
\end{bmatrix}.
\end{IEEEeqnarray}
As can be seen from~\eqref{eq:optgain}, the prescribed sparsity pattern of the feedback controller gain matrix is preserved.

\begin{table*}~\normalsize
	\centering
		\begin{IEEEeqnarray}{l}\label{eq:W}
			W^* = \IEEEnonumber\\
			\begin{bmatrix}
			\phantom{-}1.81 \times 10^{-1} & -3.19\times 10^{-1}  & \phantom{-}5.34 \times 10^{-2} & 0 & 0 & 0 & 0\\
		-3.19\times 10^{-1} &  \phantom{-} 7.06\times 10^{-1}  & -6.90\times 10^{-1}& 0 & 0 & 0 & 0\\
			\phantom{-}5.34 \times 10^{-2} & -6.90\times 10^{-1}  & \phantom{-}3.12 & 0 & 0 & 0 & 0\\
			0 & 0 & 0 & \phantom{-}4.16\times 10^{-7} & -1.62\times 10^{-5} & \phantom{-}3.54\times 10^{-5} & \phantom{-}5.60\times 10^{-5}\\
			0 & 0 & 0 & -1.62\times 10^{-5} & \phantom{-}1.00\times 10^{-3} & -2.95\times 10^{-2} & -5.87\times 10^{-3}\\
			0 & 0 & 0 & \phantom{-}3.54\times 10^{-5} & -2.95\times 10^{-2} & \phantom{-}2.74 & -3.62\times 10^{-2}\\
			0 & 0 & 0 &  \phantom{-}5.60\times 10^{-5}& -5.87\times 10^{-3} & -3.62\times 10^{-2} & \phantom{-}7.21
			\end{bmatrix}.\IEEEnonumber \\
		\end{IEEEeqnarray}
\end{table*}

\subsection{Experimental Results}

\begin{figure}[t]
	\centering
	\includegraphics[trim=50 50 100 100,width=1\columnwidth]{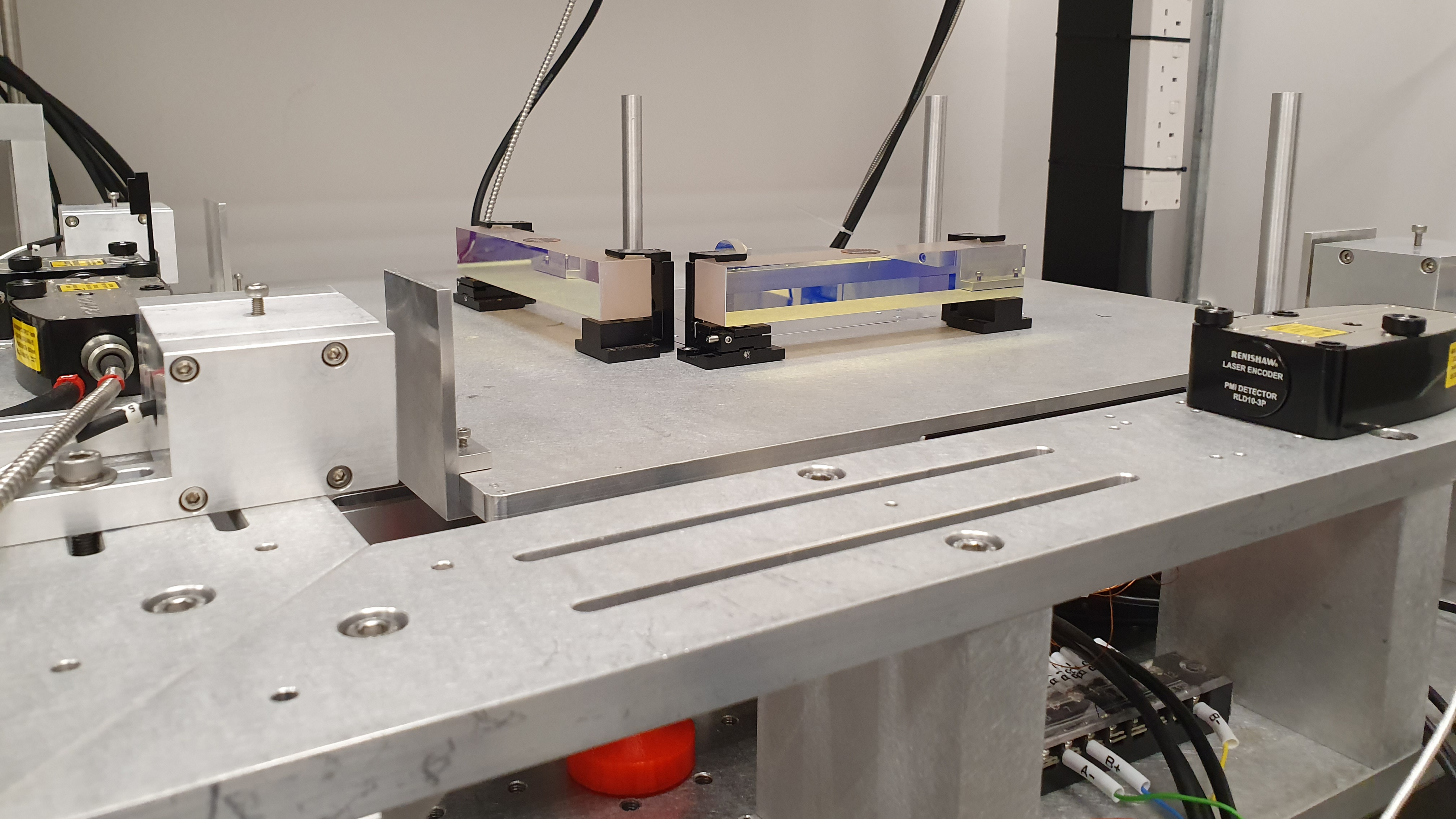}
	\caption{Prototype of the maglev planar positioning system used in this work.}
	\label{fig:setup}
\end{figure}

For control of the 6-DoF motion, three Renishaw fiber optic laser interferometers are used to measure the in-plane motions with a count resolution of 40 nm, and three Lion Precision capacitive sensors with the root mean square (RMS) resolution of 150 nm are used to measure the out-of-plane motions. Trust TA320 and TA115 linear amplifiers are used to actuate the 8-phase coils with an upper limit of 1.2 A. For all experiments conducted in this work, the levitation height is fixed at 1\,mm constantly with zero rotation angles with respect to $x-$, $y-$, and $z-$axes.

The 2-DoF controller is implemented in the maglev planar positioning system shown in Fig.~\ref{fig:setup}. As in~\eqref{eq:ff}, the feedforward control input is calculated and implemented based on the parameters of the nominal model, and the feedback controller is implemented using the optimized results. Notice that the controller is discretized in a sampling frequency of 2.5 kHz. Real-time experiments are conducted for validation of our proposed method (denoted by Method 1), where one experiment is implemented without any load, and the other experiment is implemented with a given load of 0.6 kg. Additionally, comparative experiments are carried out using the loop shaping method as proposed in~\cite{zhu2017analysis} for the similar maglev planar positioning system (denoted by Method 2).

\begin{figure}[t]
	\centering
	\includegraphics[trim=50 50 50 100,width=1\columnwidth]{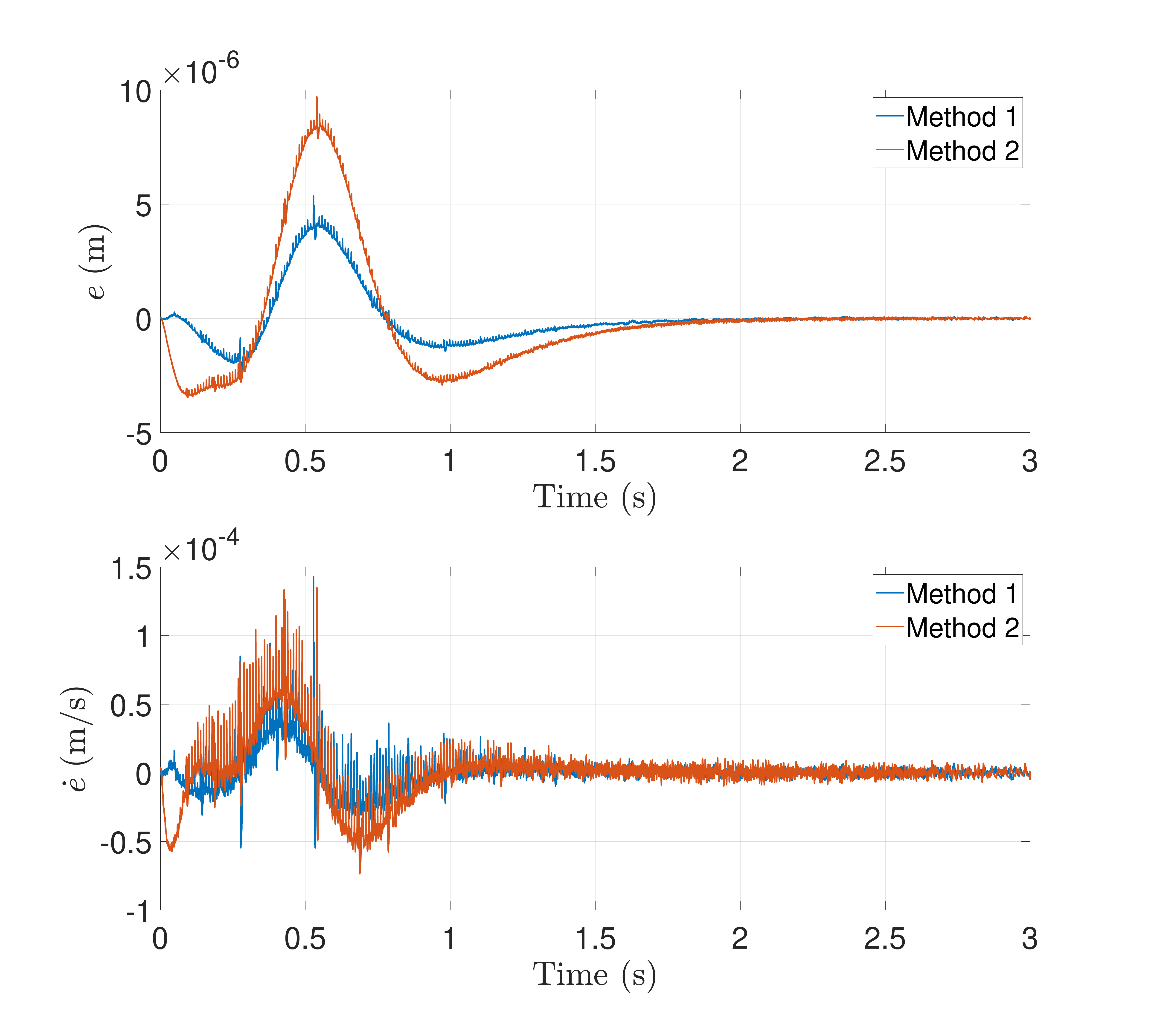}
	\caption{Position error and its derivative in the experiment without the load.}
	\label{fig:error_no_load}
\end{figure}

\begin{figure}[t]
	\centering
	\includegraphics[trim=50 50 50 100,width=1\columnwidth]{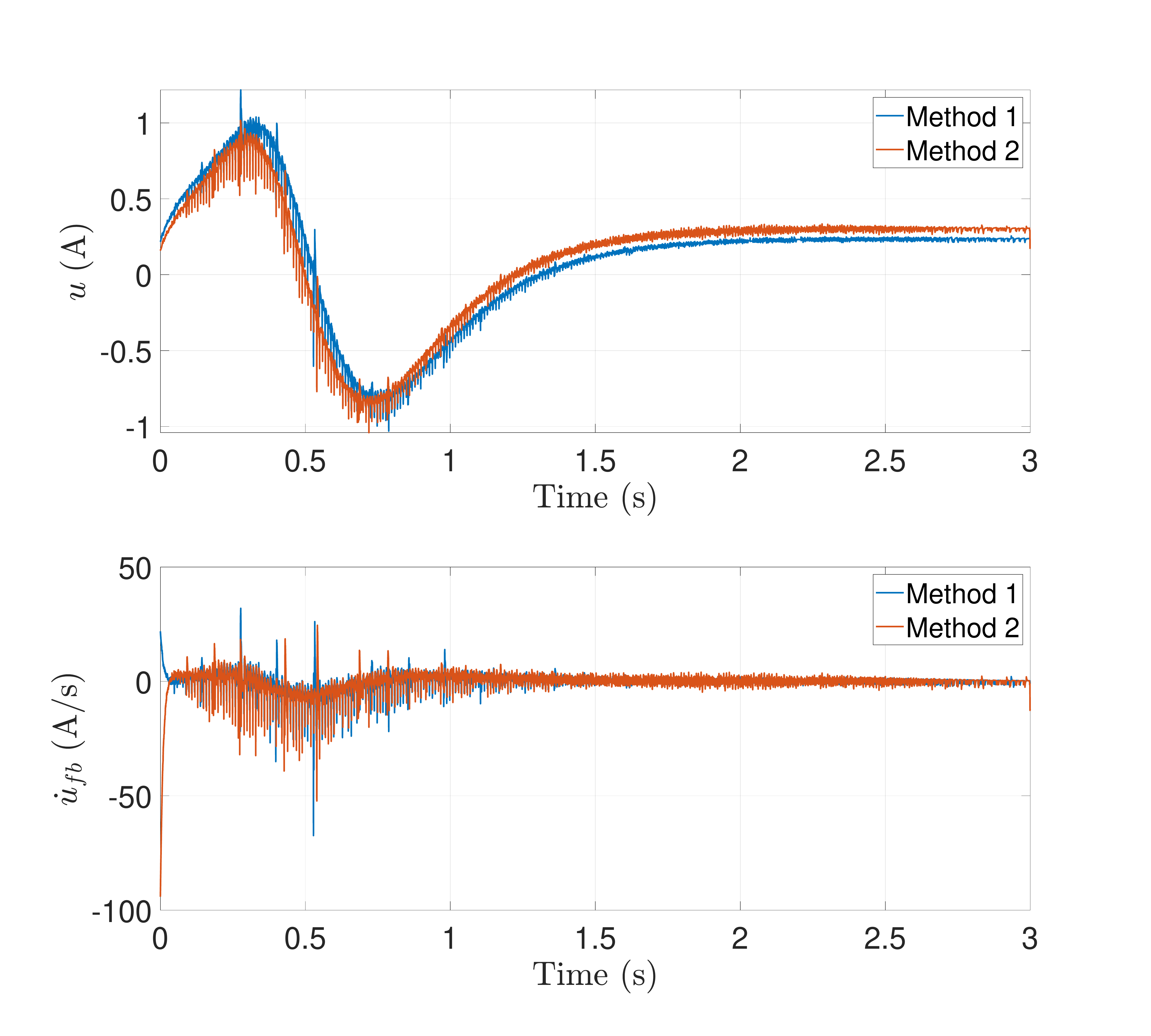}
	\caption{Control input and feedback control input chattering in the experiment without the load.}
	\label{fig:input_no_load}
\end{figure}

For the experiment without any load, Fig.~\ref{fig:error_no_load} shows the position error and its derivative, where the root-mean-square (RMS) values of them are documented in Table~\ref{table:tab}. The experimental results show that, compared with Method 2, our proposed approach gives better tracking performance in terms of precision and smoothness. Fig.~\ref{fig:input_no_load} shows the control input as well as the chattering of the feedback control input. Note that in this work, $\dot e$ and $\dot u_{fb}$ are derived by passing the signals $e$ and $u_{fb}$ through a differentiator and a low-pass filter given by $100s/(s+100)$. It is interesting to see that the control input and its chattering at $t=0$ are not zero, and this phenomenon arises from the integration of error signals before the start of the motion control experiment, because the stage needs to be floated beforehand. Additionally, as can be seen from Fig.~\ref{fig:input_no_load}, the feedback control input chattering is suppressed within an adequate level without strong vibration incurred in the experiment, and the chattering in the proposed method is less than  Method 2. Besides, the RMS values of the feedback control input chattering are given in Table~\ref{table:tab}, which further validates our proposed approach.


\begin{table}[t]\centering
	\caption{Comparative experimental results}
	\label{table:tab}
\begin{tabular}{|c|c|c|c|}
	\hline
	\multicolumn{2}{|c|}{}                              & \begin{tabular}[c]{@{}c@{}}Method 1\\ (RMS Value)\end{tabular} & \begin{tabular}[c]{@{}c@{}}Method 2\\ (RMS Value)\end{tabular} \\ \hline
	\multirow{3}{*}{Without Load} & $e$ (m)             & $1.16 \times 10^{-6}$                                           & $2.47 \times 10^{-6}$                                           \\ \cline{2-4} 
	& $\dot e$ (m/s)      & $1.17 \times 10^{-5}$                                           & $2.01 \times 10^{-5}$                                           \\ \cline{2-4} 
	& $\dot u_{fb}$ (A/s) & 3.45                                                            & 5.22                                                            \\ \hline
	\multirow{3}{*}{With Load}    & $e$ (m)             & $1.14 \times 10^{-6}$                                           & $2.62 \times 10^{-6}$                                           \\ \cline{2-4} 
	& $\dot e$ (m/s)      & $1.34 \times 10^{-5}$                                           & $2.18 \times 10^{-5}$                                           \\ \cline{2-4} 
	& $\dot u_{fb}$ (A/s) & 4.33                                                            & 5.67                                                            \\ \hline
\end{tabular}
\end{table}

\begin{figure}[t]
	\centering
	\includegraphics[trim=50 50 50 80,width=1\columnwidth]{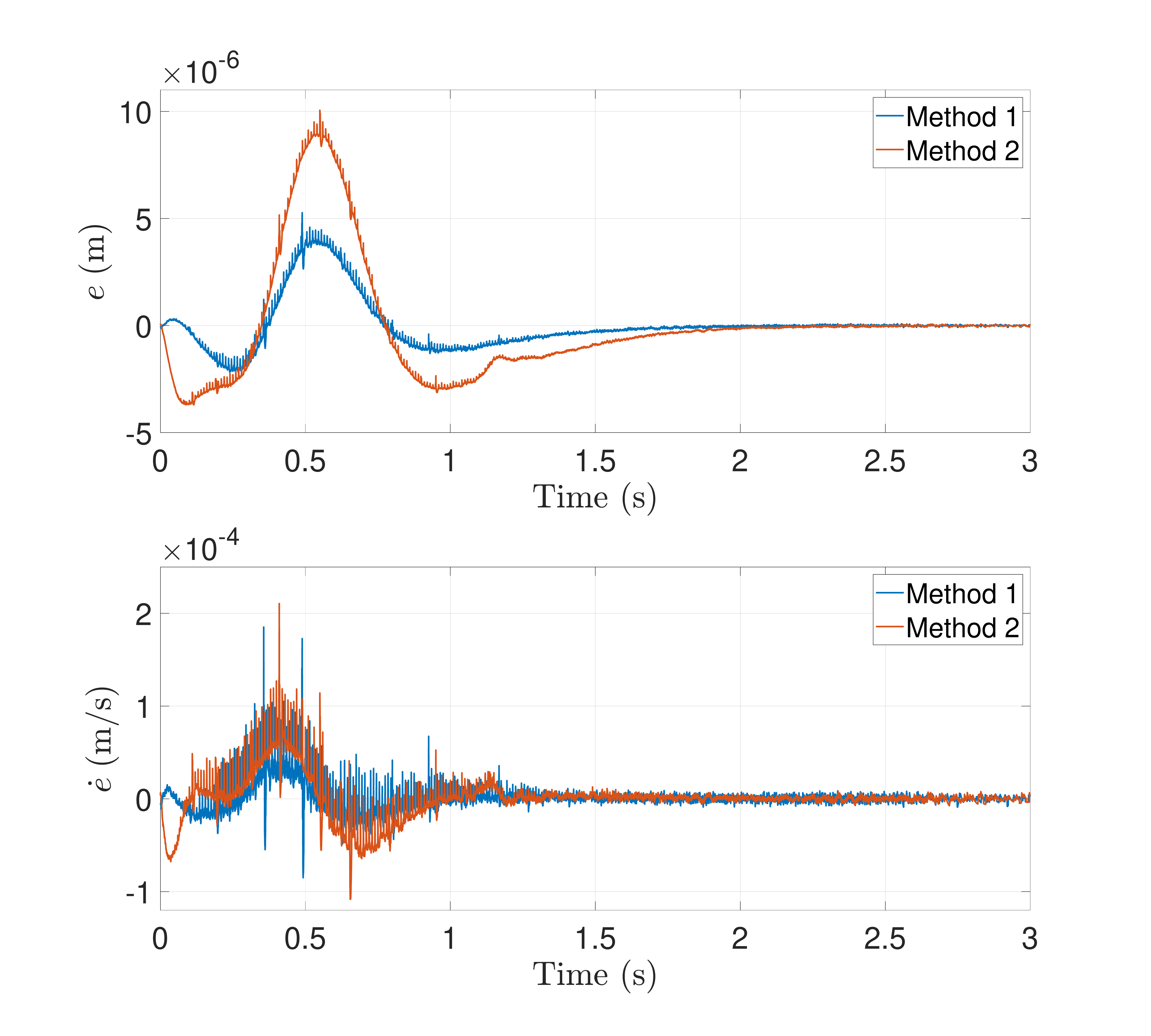}
	\caption{Position error and its derivative in the experiment with the load.}
	\label{fig:error_load}
\end{figure}

\begin{figure}[t]
	\centering
	\includegraphics[trim=50 50 50 100,width=1\columnwidth]{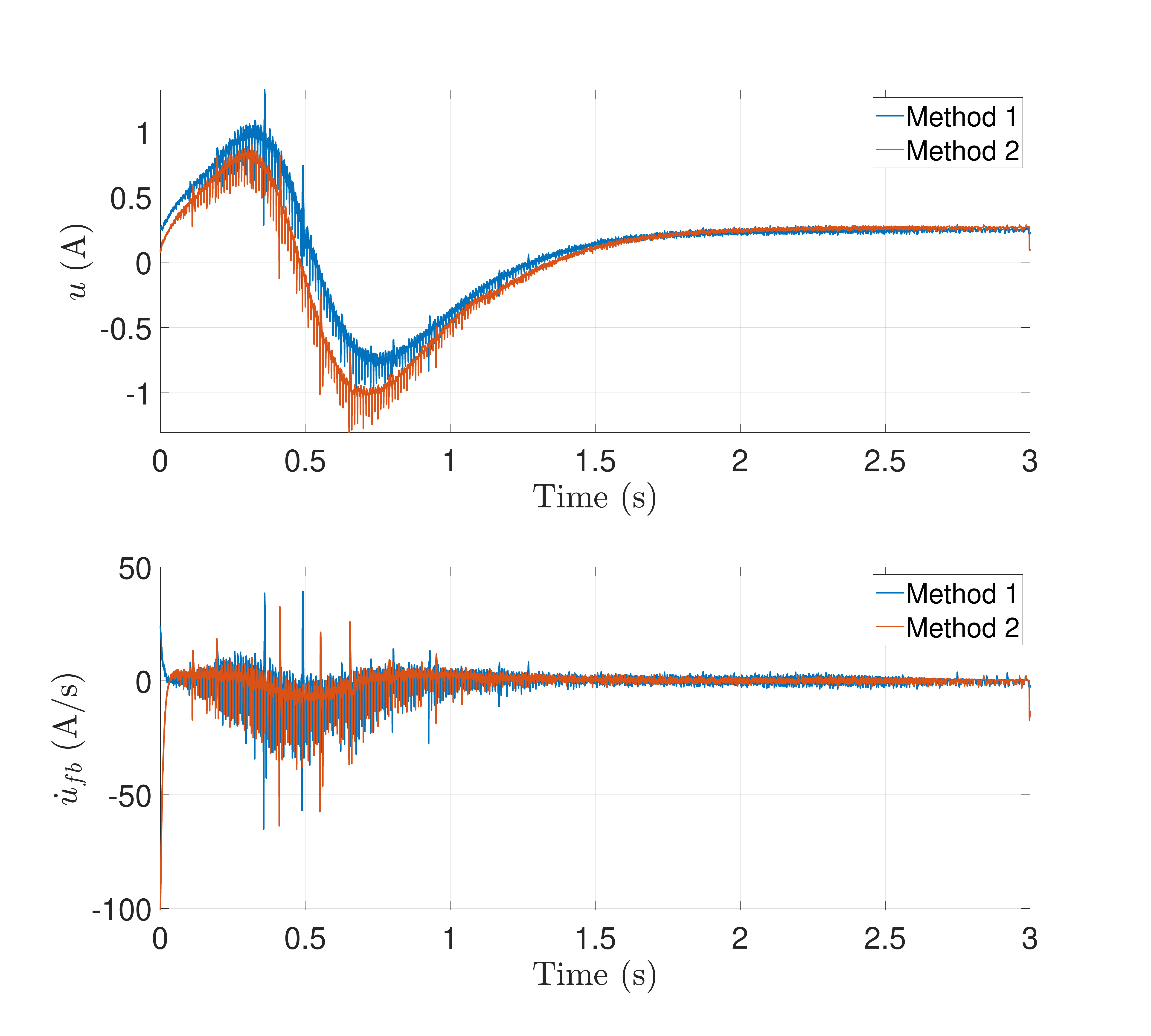}
	\caption{Control input and feedback control input chattering in the experiment with the load.}
	\label{fig:input_load}
\end{figure}

After the load is placed, the tracking performance of the system is shown in Fig.~\ref{fig:error_load}. It can be observed that the maglev planar positioning stage can still track the reference profile rather accurately and smoothly. Fig.~\ref{fig:input_load} depicts the control input and the feedback control input chattering. Besides, the RMS values of the position error, the derivative of the position error, and the feedback control input chattering are provided in Table~\ref{table:tab}. With our method, the RMS value of $e$ is decreased by 1.75\% by applying the load. On the other hand, the RMS value of $\dot e$ and $\dot u_{fb}$ is increased by 14.53\% and 25.51\%, respectively. With Method 2, the RMS value of $e$, $\dot e$ and $\dot u_{fb}$ is increased by 6.07\%, 8.46\%, and 8.62\%, respectively, by applying the load. Thus it can be clearly observed the robustness of our proposed approach. Based on these data, we think our improvement is significant enough in this maglev application. It can be clearly seen that, with the proposed method, the effect caused by the applied loading is less significant than the method adopted in~\cite{zhu2017analysis}.


\begin{figure}[t]
	\centering
	\includegraphics[trim=50 50 50 80,width=1\columnwidth]{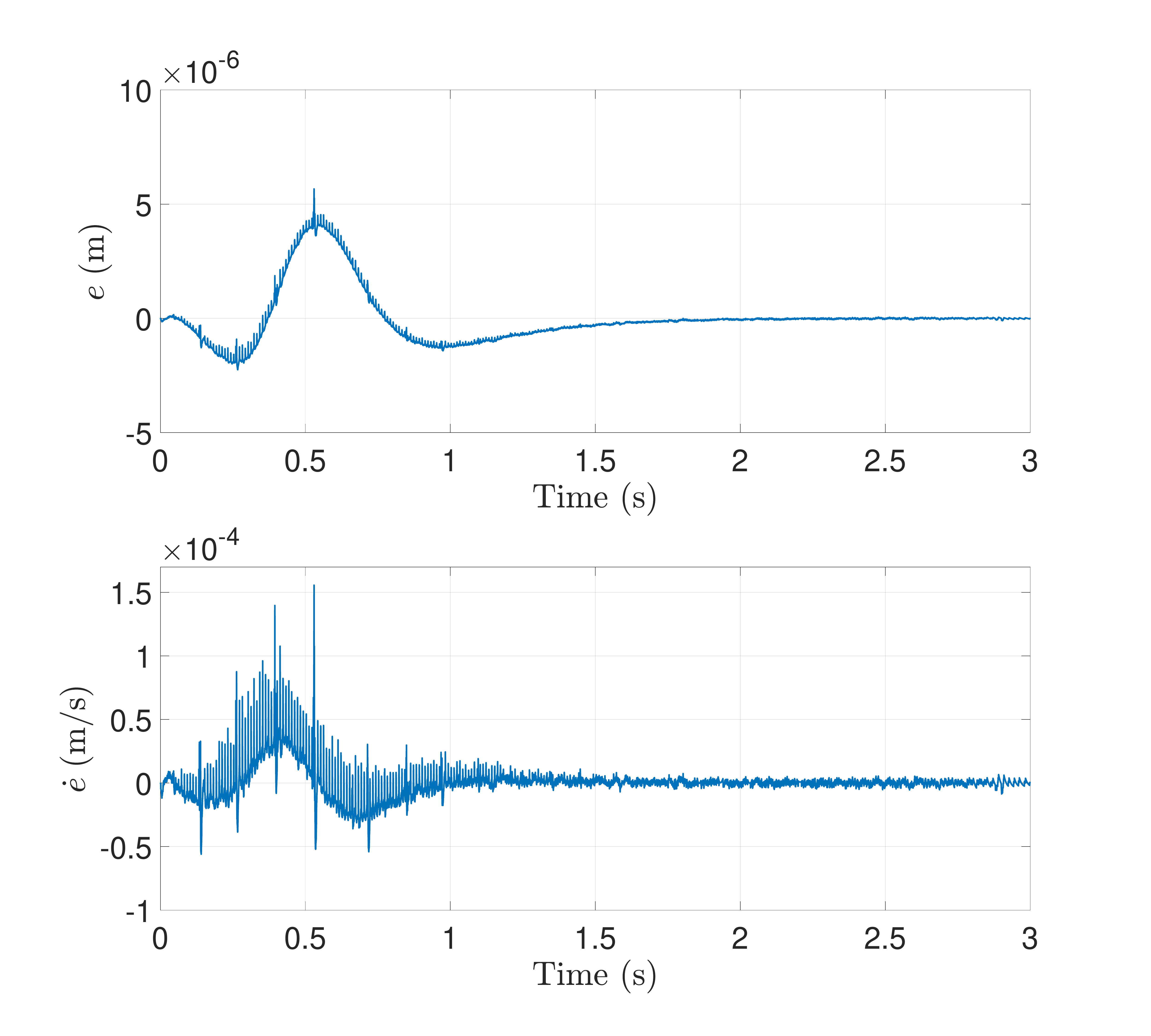}
	\caption{Position error and its derivative in the experiment with noises.}
	\label{fig:error_noload_noise}
\end{figure}

\begin{figure}[t]
	\centering
	\includegraphics[trim=50 50 50 100,width=1\columnwidth]{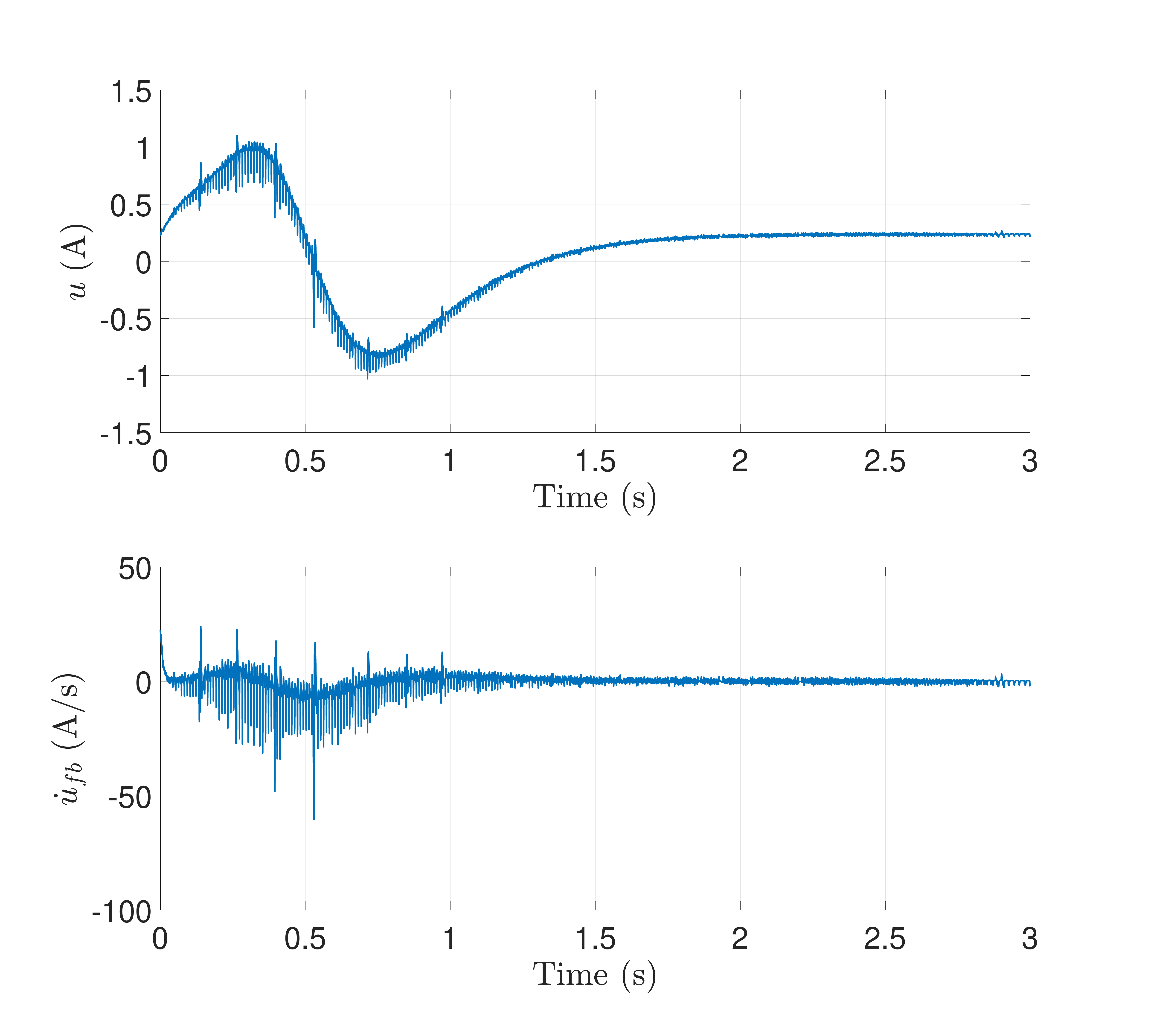}
	\caption{Control input and feedback control input chattering in the experiment with noises.}
	\label{fig:input_noload_noise}
\end{figure}

Additionally, to test the disturbance rejection capability of the proposed method, noises with a uniformly distribution in the range [-0.05 N, 0.05 N] are added to the decoupled axis. The position error of the system and its derivative are shown in Fig.~\ref{fig:error_noload_noise}. Also, Fig.~\ref{fig:input_noload_noise} shows the control input and the feedback control input chattering. In this experiment, the RMS value of $e$, $\dot e$, and $\dot u_{fb}$ are given by $1.17 \times 10^{-6}$ $\textup{m}$, $1.22 \times 10^{-5}$ $\textup{m}/\textup{s}$, and $3.76 $ $\textup{A}/\textup{s}$, respectively. Compared with the results attained without the addition of noises, the performance is slightly worse, but the maglev planar positioning stage still gives satisfying performance despite the existence of noises. Hence, the disturbance rejection capability of the proposed method is successfully validated.


To conclude, the experimental results clearly show that the proposed controller design approach is able to accommodate the parametric uncertainties and external noises in the maglev planar positioning stage, and attain good robust stability and system performance such that an accurate and smooth tracking task is successfully achieved.

\section{Conclusion}

In this paper, the accurate and smooth tracking problem of a maglev planar positioning stage is presented with a pre-defined S-curve reference trajectory. First, the feedforward component of the controller is designed according to the S-curve reference trajectory and the nominal model of the maglev system. Second, through the ${H}_\infty$ control formulation and the invoked convex parameterization, a set of robust stabilizing feedback controllers with the prescribed sparsity pattern are rather elegantly parameterized over a convex set, and thus a convex optimization problem is formulated. The optimization problem is efficiently solved and a global optimal solution of the feedback controller is obtained. With the calculated parameters, the 2-DoF controller is implemented. The experimental results on the maglev planar positioning stage successfully validate the effectiveness and practicability of the proposed methodology, where the stringent requirements on robust stability and system performance are all very well achieved despite the existence of parametric uncertainties.

\bibliographystyle{IEEEtran}
\bibliography{IEEEabrv,Reference}

\end{document}